\documentclass[12pt]{iopart}

\usepackage{graphicx}
\usepackage{iopams}

\newcommand\A{{\scriptscriptstyle \rm ANNNI}}

\begin{document}

\title{Strongly interacting Majorana modes in an array of Josephson
junctions}

\author{Fabian Hassler$^1$ and Dirk Schuricht$^{2,3}$}
\address{$^1$ Institute for Quantum Information, RWTH Aachen University, 
52056 Aachen, Germany}
\address{$^2$ Institute for Theory of Statistical Physics, 
RWTH Aachen University, 52056 Aachen, Germany}
\address{$^3$ JARA-Fundamentals of Future Information Technology}
\ead{hassler@physik.rwth-aachen.de, schuricht@physik.rwth-aachen.de}
\pacno{
 75.10.Pq,  
 71.27.+a,  
 74.81.Fa,  
 73.23.Hk  
}

\begin{abstract} An array of superconducting islands with semiconducting
   nanowires in the right regime provides a macroscopic implementation of
   Kitaev's toy model for Majorana wires. We show that a capacitive coupling
   between adjacent islands leads to an effective interaction between
   the Majorana modes. We demonstrate that even though strong repulsive
   interaction eventually drive the system into a Mott insulating state
   the competition between the (trivial) band-insulator and the (trivial)
   Mott insulator leads to an interjacent topological insulating state for
   arbitrary strong interactions.
\end{abstract}

\maketitle

\section{Introduction}

Majorana zero modes are fermions which are their
own antiparticles. They are believed to exist as effective particles
in the middle of the gap of a topological superconductor. The recent
interest in Majorana zero modes originates in their non-Abelian exchange
statistics~\cite{moore:91}, which is the basis for potential applications
in quantum computation~\cite{kitaev:03}.  Based on the theoretical
proposal~\cite{wire} to realize this exotic states in semiconducting
nanowires with strong spin-orbit coupling in a magnetic field and in
proximity to a conventional (nontopological) superconductor, recent
experimental progress has shown signatures of Majorana zero modes in the
tunnelling conductance of normal conducting-superconducting~\cite{mourik:12} 
and superconducting-normal conducting-superconducting systems~\cite{deng:12}.
More information and references on the fast developing field can be found
in the recent reviews~\cite{beenakker:11}.

As the nanowire is one-dimensional interaction effects become important.
On the one hand, employing field theoretical methods Gangadharaiah \emph{et
al.}~\cite{gangadharaiah:11} have argued that strong electron-electron
interactions generically destroy the topological phase by suppressing the
superconducting gap. On the other hand, using a combination of analytical
and numerical methods Stoudenmire \emph{et al.}~\cite{stoudenmire:11}
have shown that repulsive interactions significantly decrease the
required Zeeman energy and increase the parameter range for which
the topological phase exists.  It is believed that the origin of
these effects is an interaction driven renormalization of the Zeeman
gap~\cite{stoudenmire:11,braunecker:10}. Furthermore, it has been shown that
in helical liquids the scattering processes between the constituent fermion
bands open gaps which in turn lead to a stabilization of the Majorana states
against interactions~\cite{sela:11} and that (an odd number of) Majorana zero
modes are in fact stable against general interactions~\cite{goldstein:11}.
For further studies of the effect of electron-electron interactions on Majorana zero 
modes in nanowires and two-chain ladders see~\cite{interactions}.

Recently, a macroscopic version of the Kitaev chain \cite{kitaev:01},
a toy model for a nanowire supporting noninteracting Majorana modes, has
been proposed \cite{heck:11} in a one-dimensional (1D) array of topological
superconducting islands. Its advantage over microscopic implementations is
that the individual parameters of the effective Hamiltonian can potentially
be tuned in situ.  Here, we show that additional capacitances between
adjacent islands lead to an effective interaction between the low-energy
Majorana degrees of freedom.  We present the phase diagram of the system
which demonstrates that sufficiently strong repulsive interactions will
drive the system from the topologically trivial phase into the topological
phase supporting Majorana zero modes before eventually leading to a Mott
insulating state. We discuss how the parameters of the system can be tuned
by changing the gate voltages and exploit this to propose the detection
of the different phases and phase boundaries in a tunnelling experiment.

\begin{figure}[t]
	\centering
	\includegraphics[width=0.7\textwidth]{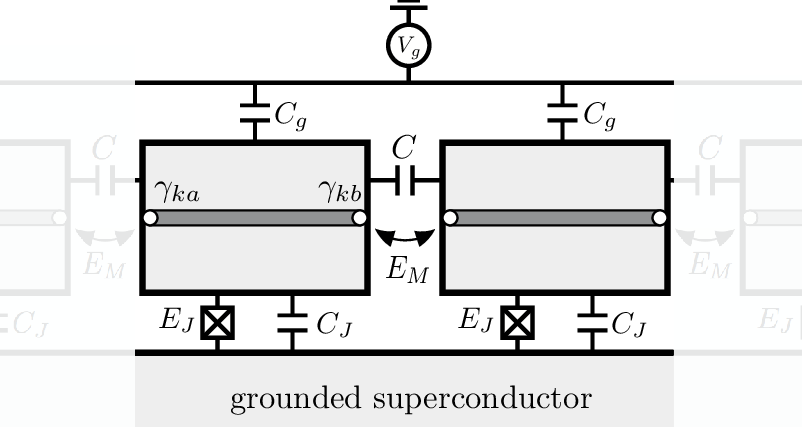}
	\caption{A 1D array of superconducting islands (light
  grey) coupled via strong Josephson junctions ($E_J$) to a common ground
  superconductor. Each island contains a pair of Majorana zero modes (white
  dots) at the end points of a semiconducting nanowire (dark grey). The tunnel
  coupling of individual electrons between the superconducting islands
  is proportional to the energy scale $E_M$. A common gate voltage $V_g$
  can be used to tune the relative strength of the different terms in the
  Hamiltonian. The capacitive couplings between the elements are denoted
  by $C$, $C_J$, and $C_g$, respectively.}
  \label{fig:model}
\end{figure}

\section{Model}

We discuss a system consisting of a
1D array of $N$ superconducting islands (see figure~\ref{fig:model}).
Because of the proximity-coupled semiconducting nanowire each of the
islands has two midgap Andreev states, i.e., Majorana modes, located at
the ends of the nanowire \cite{wire}.  We will denote with $\gamma_{ka}$
and $\gamma_{kb}$ the two Majorana operators on island $k$ associated
with these zero modes.  The Majorana operators are Hermitian $\gamma_\mu =
\gamma_\mu^\dag$ and fulfil the Clifford algebra $\{\gamma_\mu, \gamma_\nu
\} = 2 \delta_{\mu\nu}$. The total Lagrangian $L= T - V_J - V_M$ of the
system consists of three terms. Coupling of the Majorana modes on nearby
islands leads to the term~\cite{kitaev:01,fu:10} $
  V_M = i\,E_M \sum_{k=1}^N \gamma_{kb} \gamma_{k+1a} \cos [(\phi_{k+1}
  - \phi_k)/2]$
where $\phi_k$ is the superconducting phase of the $k$-th island. Here and
in the following, we assume for convenience periodic boundary condition such
that islands 1 and $N+1$ are equivalent.  Apart from the Majorana modes,
the term discussed above has the additional degrees of freedom $\phi_k$
due to the condensate of Cooper pairs.  Similar to \cite{heck:11},
we eliminate these by connecting each superconducting island with a strong
Josephson junction to a common (ground) superconductor.  This fixes the
superconducting phases (up to some quantum phase-slips discussed below) and
is described by the Hamiltonian $V_J = E_J \sum_{k=1}^N (1- \cos\phi_k);$
here, $E_J = \hbar I_c/2e$ is the effective Josephson coupling of each
of the Josephson junctions with critical current $I_c$. In the limit $E_J
\gg E_M$ the junctions effectively pin the phases of all superconducting
islands to a common value $\phi_k \equiv 0$.  As a result $V_M$ reduces
to the pure Majorana coupling
\begin{equation}\label{eq:ham_m}
  H_M = i\,E_M \sum_{k=1}^N \gamma_{kb} \gamma_{k+1a}, \qquad (E_J \gg E_M).
\end{equation}
Finally, the kinetic term $T = (\hbar^2/8e^2) \sum_{k=1}^N [ C (\dot
\phi_{k+1} - \dot \phi_k)^2 + C_G \dot \phi_k^2 ] + (\hbar/2e) \sum_{k=1}^N
q \dot\phi_k $ occurs due to the capacitive couplings, where $C$ denotes
the capacitance between neighbouring islands and $C_G =C_g + C_J$ the
(total) capacitance to the ground. Here, $C_J$ is the capacitance of the
strong Josephson junction and $C_g$ the capacitance to a common back gate
at voltage $V_g$ with respect to the ground superconductor. Apart from
the charging energy, the back-gate introduces a term proportional to
the induced charge $q = C_g V_g$ which is tunable with single-electron
precision~\cite{tune} via $V_g$ and whose effect will become important
later. The typical capacitive energy scale $E_C= e^2/2C_\Sigma$ of a single
islands depends on the total capacitance $C_\Sigma = 2 C + C_G$.

\section{Mapping on an effective spin model}

As we have seen the strong coupling to the
ground superconductor pins the superconducting phase differences to $\phi_k
\equiv 0$ and thus changes the energy due to the Majorana modes from $V_M$
to $H_M$. The effect of the strong coupling to the ground superconductor
on the charging energy $T$ is more subtle. We first present the results for
$C=0$ before extending them to nonzero $C$: as charge and phase are conjugate
variables the pinning of the phases $\phi_k$ greatly reduces the effect of
charging \cite{heck:11}. An effective charging energy still arises due to
quantum phase slips through the junctions.  For example, changing $\phi_k$
from $0$ to $2\pi$ leads to a charging energy ($E_J \gg E_C$) \cite{koch:07}
\begin{equation}\label{eq:ham_c}
  H_{Ck} = 
  \Gamma_\Delta 
  \cos[\pi(q/e + n_k)]=  \Gamma_\Delta \cos(\pi q/e ) \mathcal{P}_k.
\end{equation}
Here, the tunnelling amplitude is given by $\Gamma_\Delta \simeq E_C^{1/4}
E_J^{3/4} e^{-S_\Delta}$ with $S_\Delta= \hbar^{-1} \int\!d\tau\,L_E
=\sqrt{8E_J/E_C}$ the dimensionless Euclidean action along the classical
trajectory $\gamma_\Delta\colon \phi_k \in [0,2\pi]$ in the inverted
potential with $L_E(\tau) = -L(t=i\tau)$.\footnote{Because $V_J \gg
V_M$, we take only the potential $V_J$ into account when calculating the
tunnelling action $S_E$.} The cosine term in \eref{eq:ham_c} occurs due to the
Aharonov-Casher interference of the two tunnelling paths $\phi_k=0 \to 2\pi$
and $\phi_k=0 \to - 2\pi$ which lead to an indistinguishable final state
and thus interfere with a phase difference depending on the total induced
charge $q+ e n_k$ \cite{ivanov:02,heck:11}.  The term $n_k = \frac12 (1-
\mathcal{P}_k)\in \{0,1\}$ is the contribution to the charge due to the
parity $\mathcal{P}_k = i \gamma_{ka} \gamma_{kb} \in \{-1,1\}$ of the
number of electrons on the superconducting island encoded in the state of
the Majorana zero modes~\cite{fu:10,heck:11}. Equation \eref{eq:ham_c} is a
chemical potential term: For large $\Gamma_\Delta$, all the fermionic states
are either filled or empty and the system is in a band insulating state.

Going away from this special point and introducing a finite cross-capacitance
parametrized by $\eta = 2 C/C_\Sigma$ with $\eta \in [0,1]$, the classical
path for a phase slip in $\phi_k$ does not only involve $\phi_k$ but also
the other phases. To lowest nonvanishing order in $\eta$, we only need to
take into account the change in $\phi_{k-1}$ and $\phi_{k+1}$ and obtain
$S_\Delta = \sqrt{8 E_J/E_C} [1 + (\pi^2 -12)\eta^2/96]$, which is accurate
to more than 2 digits all the way up to $\eta=1$ as numerics confirms.
Additionally, the effective capacitance coupling between two islands becomes
important. This term is generated by simultaneous phase slips of the two
phases $\phi_k$ and $\phi_{k+1}$ of neighbouring islands. Due to the symmetry
between island $k$ and $k+1$, we have $\phi_k=\phi_{k+1}$ for the classical
path $\gamma_U\colon\phi_k = \phi_{k+1} \in [0,2\pi]$ with the corresponding
Euclidean action $S_U= \sqrt{16 (2- \eta) E_J/E_C}$.  The Aharonov-Casher
interference in this case leads to the tunnelling amplitude $\Gamma_U \simeq
E_C^{1/4} E_J^{3/4} e^{-S_U}$ depending on the total charge $2 q +e (n_k
+n_{k+1})$ of the two islands involved. Thus we obtain the interaction term
\begin{eqnarray}\label{eq:ham_u}
  H_{Uk} = \Gamma_U \cos[2\pi q/e + \pi (n_k + n_{k+1})] 
  =  
  \Gamma_U \cos(2\pi q /e) \mathcal{P}_k \mathcal{P}_{k+1}.
\end{eqnarray}
We stress that \eref{eq:ham_u} is an interaction term involving
four Majorana operators and that the amplitude is modulated by twice
the induced charge $q$ compared to \eref{eq:ham_c}. We call $U>0$
repulsive interaction as it prefers having an occupied site next to an
empty one. For strong repulsive interactions the system is driven into a
Mott insulator or equivalently commensurate charge density wave (CDW) state.

The complete low-energy Hamiltonian in the limit $E_J \gg E_M,E_C$ is given
by $H_\A = \sum_{k=1}^N (H_{Ck} + H_{Uk}) + H_M$, which constitutes the
transverse axial next-nearest-neighbour Ising (ANNNI) model~\cite{ANNNI} as
can be seen by performing a Jordan-Wigner transformation $\mathcal{P}_k =
i \gamma_{ka} \gamma_{kb} = \sigma^z_k$, $i \gamma_{kb} \gamma_{k+1a} =
\sigma^x_k \sigma^x_{k+1}$, resulting in the spin Hamiltonian
\begin{equation}\label{eq:annni}
  H_\A =  \sum_{k=1}^N ( \Delta \sigma^z_k 
  + U \sigma^z_k \sigma^z_{k+1} + E_M \sigma^x_k \sigma^x_{k+1}).
\end{equation}
Here, the $\sigma_k^{x,y,z}$ are Pauli matrices, $\Delta = \Gamma_\Delta
\cos(\pi q/e)$ and $U= \Gamma_U \cos(2\pi q/e) $. We note that the
energies $\Delta$ and $U$ can be tuned via the charge $q$ induced on the
superconducting islands through the gate voltage $V_g$. As the fermionic
Hamiltonian without the interaction term proportional to $U$ has been
intensively studied before~\cite{kitaev:01}, we focus here on the effects
of the interaction term. In our set-up, this term is most important in the
case $C\gg C_G$ where $\eta \approx 1$ and $\Gamma_U \approx\Gamma_\Delta
\exp[-1.25 \sqrt{E_J/E_C}]$, which can be as large as $0.3$ for $E_J
\approx E_C$; note that in this regime the actions $S_\Delta$ and $S_U$
are still much larger than one such that the semiclassical approximation
employed above is valid.

\section{Phase diagram}

We first note that the system \eref{eq:annni} is invariant
under $E_M\to -E_M$ as well as $\Delta\to-\Delta$ due to the
transformations $\sigma_k^{x,y}\to (-1)^k\sigma_k^{x,y}$ and
$\sigma_k^{x,z}\to-\sigma_k^{x,z}$ respectively. Thus without
loss of generality, we assume in the following $E_M,\Delta>0$.
The phase diagram of the spin model \eref{eq:annni} contains four
phases:\footnote{The Hamiltonian \eref{eq:annni} is brought to standard form
by performing the duality transformation $\mu_k^x=\prod_{j<k}\sigma_j^z$,
$\mu_k^z=\sigma_k^x\sigma_{k+1}^x$. The phase diagram of the resulting
model including expressions for the phase transitions has been worked
out in \cite{ANNNI,ANNNI2} in the parameters $\kappa=U/\Delta$ and
$E_M/\Delta$.} A paramagnetic phase (PM) with a unique ground state
with $\langle\sigma_k^z\rangle<0$, $\langle\sigma_k^x\rangle=0$; an
antiferromagnetic phase (AFM) with doubly degenerate ground state and $\langle\sigma_k^x\rangle\propto
(-1)^k$; an ``anti phase" (AP) with a doubly degenerate ground state with
$\langle\sigma_k^z\rangle\propto (-1)^k$; and a ``floating phase" (FP)
between the AFM and the AP. For $\Delta=0$, the duality transform of the
model \eref{eq:annni} is a sum of two quantum Ising chains, while for the
noninteracting case it reduces to a single quantum Ising chain.

\begin{figure}[t] \centering
	\includegraphics[width=0.7\textwidth,clip=true]{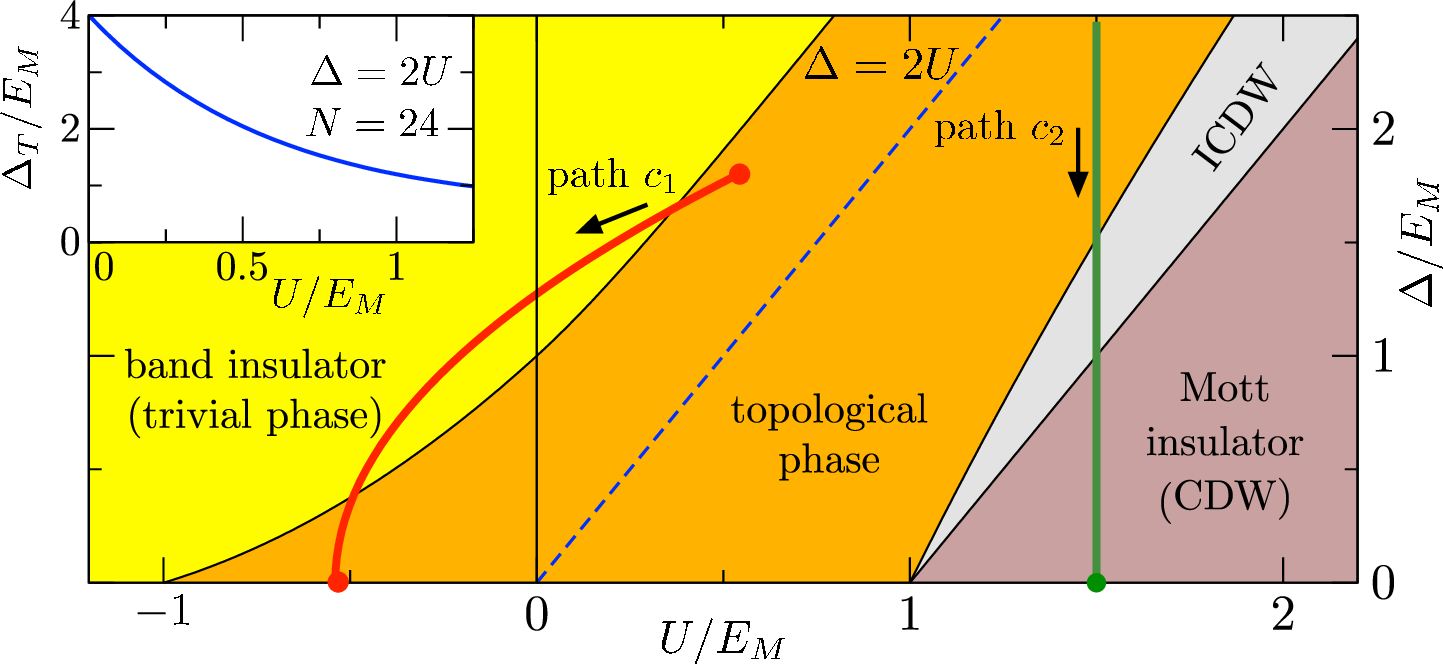}
	\caption{\label{fig:phasediagram}Phase diagram
	of the Josephson junction array \eref{eq:annni}. The phase
	diagram is invariant under $\Delta\to -\Delta$. The topological
	phase is characterized by a doubly degenerate ground state. In
	figure~\ref{fig:conductance} we show the tunnelling conductance along
	the paths $c_1$ and $c_2$. Inset: Gap $\Delta_T$ between the (nearly)
	degenerate ground states ($\delta \lesssim 0.02\,E_M$) and the
	first excited state along the blue dashed line as determined by
	exact diagonalization.}
\end{figure} 
As we have shown above in our realization of the ANNNI model the
parameters $\Delta$ and $U$ can easily be tuned via a gate voltage. On
the other hand, the coupling $E_M$ is determined by the overlaps of
the Majorana wave functions on neighbouring islands and thus fixed by
the geometry of the array. Hence it is natural to consider the phase
diagram as function of $\Delta/E_M$ and $U/E_M$, which is shown in
figure~\ref{fig:phasediagram}. For fixed $\Delta/E_M> 1$ and sufficiently
small interaction $U$ the system is in a trivial (band-insulating) phase
corresponding to the PM in the effective spin model, which is characterized
by an unique ground state with $\langle\mathcal{P}_k\rangle<0$. By
increasing $U$ we cross into a topological phase (corresponding to the
AFM) with $\langle\mathcal{P}_k\rangle=0$ and two degenerate ground
states $|\pm\rangle$, distinguished by the (total) fermion parity
$\prod_{k}\mathcal{P}_k |\pm\rangle = \pm |\pm\rangle$.  The parity
protection (also called topological protection) originates from the fact
that any fermionic perturbation conserves the fermion parity and thus
cannot mix the states $|\pm\rangle$. The phase transition between the
trivial and topological phase is, for $U>0$, located at
\begin{equation}
  1-\frac{2U}\Delta = \frac{E_M}\Delta - \frac{U E_M^2}{2 \Delta^2 (\Delta
  -U)}, \quad \frac{E_M}{\Delta}\ll 1.
\end{equation}
In particular, we find $\Delta=2U$ for $U\gg E_M$. On the other hand, for $|U|\ll E_M$ perturbation theory yields $\Delta=E_M+32\,U/(6\pi)$.

At large positive $U$ we eventually enter incommensurate and commensurate
CDW (Mott insulator) states corresponding to the FP and the AP,
respectively.\footnote{We note that the existence of the FP, and thus
the incommensurate CDW state, at small $\Delta/E_M$ has not yet been
fully established \cite{ANNNI2}.} This region of the phase diagram
cannot be reached as long as the induced charge $q$ on the islands is
homogeneous. However, replacing the common back gate by individual gates
for each island yields the model \eref{eq:annni} with site-dependent
parameters $\Delta_i=\Gamma_\Delta\cos(\pi q_i/e)$ and $U_i= \Gamma_U
\cos(\pi(q_i+q_{i+1})/e)$ where $q_i=G_gV_g^i$ denotes the induced charge
on island $i$. Now using $q_i=(-1)^iq$ one can enter the Mott phase for
$q\to e/2$. Note that the two degenerate ground states in the Mott insulator
phase have the same fermion parity and thus are not parity protected.

The main feature of the phase diagram is its strong anisotropy under $U\to
-U$.  While negative interactions suppress the topological phase, for $U>0$
the ordering tendencies of the first and second term in \eref{eq:annni}
compete with each other.  In particular, starting from a noninteracting
point in the trivial phase, i.e., $U=0$ and $\Delta>E_M$, competition
between the band- and the Mott-insulator will drive the system into the
topological phase irrespective of the value of $\Delta/E_M$.

As discussed above the topological phase is characterized by the existence of
a doubly degenerate ground state with different fermion parity. In a finite
system this degeneracy is lifted and the value of the resulting gap $\delta$
depends on the number of islands $N$ as well as the system parameters
$\Delta/E_M$ and $U/E_M$. On the other hand, the existence of Majorana end
modes is protected by the gap between the (nearly) degenerate ground states
and the first excited state, which we denote by $\Delta_T$.  This gap is
given by $\Delta_T=4E_M$ at $\Delta=U=0$ and decreases when going away from
this point. However, as we show in the inset in figure~\ref{fig:phasediagram},
it remains of the order of $E_M$ and significantly larger than $\delta$
for $U\neq 0$ as long as one stays away from the phase boundaries.

\section{Tunnelling conductance}

After presenting the phase diagram, we
now turn to its experimental signatures in the tunnelling conductance.
The parameters $\Delta$ and $U$ can be directly tuned through the
induced charge $q$ on the islands via a common back gate voltage
$V_g$ making it possible to choose the parameters such that the path
crosses one or more phase boundaries.  

\begin{figure}[t]
	\centering
	\includegraphics[width=0.7\textwidth,clip=true]{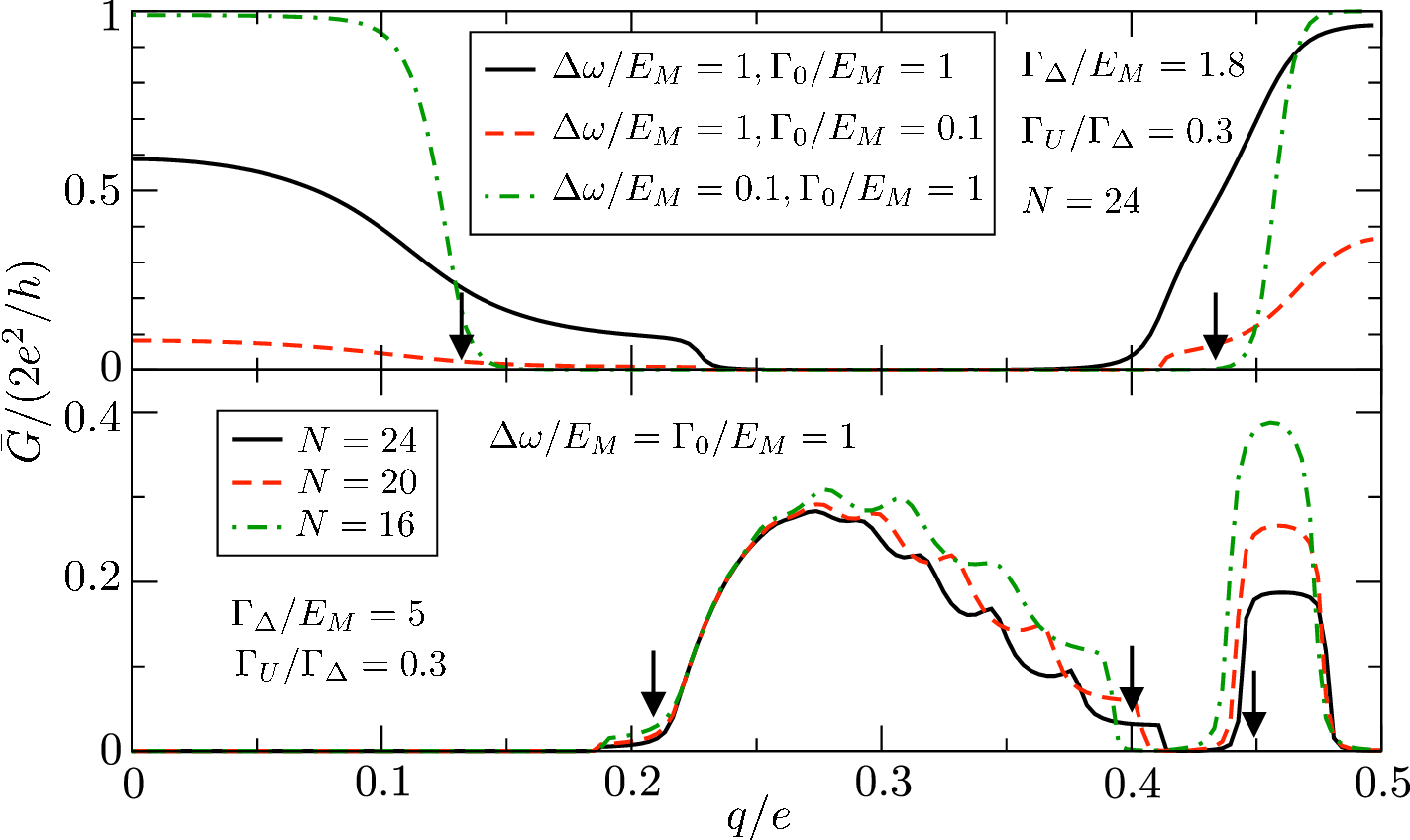}
	\caption{\label{fig:conductance}Tunnelling conductance along the paths 
	(a) $c_1$ and (b) $c_2$ shown in figure~\ref{fig:phasediagram}. 
	The arrows indicate the approximate phase boundaries.}
\end{figure}

Specifically, we consider the path $c_1(q)=(\Gamma_U\,\cos(2\pi q/e),
\Gamma_\Delta\,\cos(\pi q/e))$ with $\Gamma_U=0.3\,\Gamma_\Delta=0.54\,E_M$
such that the starting and end points at $q=0$ and $q=e/2$ lie in the
topological phase while the path enters the trivial phase in between
(see the red curve in figure~\ref{fig:phasediagram}). In the following
we consider an open chain of $N$ islands. Following \cite{law:09},
we couple the system to an electronic lead with the tunnel Hamiltonian
$H_T = t \gamma_{1a}\sum_{k\sigma} (c_{k\sigma} - c^\dag_{k\sigma})$,
where $c_{k\sigma}$ are the annihilation operators of the electrons in the
lead and $t$ is the tunnelling amplitude. We assume a constant density of
states $\rho_0$ in the lead such that the (bare) tunnelling probability is
given by $\Gamma_0 = 2\pi t^2 \rho_0$. We have calculated the differential
Andreev conductance $G(V)$ using exact diagonalization for an open chain
of length $N=24$ taking only the two lowest energy states (with different
fermion parity) into account.\footnote{The two-level approximation is
appropriate in the topological phase (where we have two levels separated
by $\Delta_T\gg\delta$ from the rest), as well as in the trivial phase
(where there is a unique ground state) and in the Mott phase (where there
are two degenerate ground state with the same fermion parity) as $\delta
\gtrsim \Delta \omega$ in the latter cases such that the current vanishes.}
In figure~\ref{fig:conductance}(a), the broadened conductance
\begin{equation} \bar{G}=
\int_{-\Delta\omega/2}^{\Delta\omega/2}\frac{d(eV)}{\Delta\omega}G(V)
\end{equation}
is plotted along $c_1$; the broadening $\Delta \omega \simeq \max(eV,k_B
T)$ is given by the maximum of bias voltage $V$ and temperature $T$.
For strong coupling $\Gamma_0$ to the lead as well as small broadening
$\Delta\omega$, the phase boundaries are clearly visible as points where
the conductance jumps from one to zero and vice versa. Weaker coupling to
the lead will lead to a suppression of the conductance in the topological
phase, i.e., unitary conductance cannot be observed. On the other hand, a
larger broadening will eventually smear out all transitions. Thus in order
to enable an experimental detection of the phase boundaries the energy
broadening should be small (in units of $E_M$) while the coupling to the
lead has to be sufficiently strong.  Recent experiments on proximity coupled
nanowires \cite{mourik:12} indicate that a Majorana coupling $E_M/k_B \simeq
100\,$mK is realistic, which is well in the range of experimental accessible
temperatures. The Majorana coupling sets the topological gap $\Delta_T$
(see inset of figure~\ref{fig:phasediagram}) as the other couplings ($E_J$
and $E_C$) can be designed in a large parameter range \cite{devoret:04}.

In order to study the regime of strong interactions we use the set-up
with individual back gates and induced charges $q_i=(-1)^iq$. In this
set-up we consider the path $c_2(q)=(\Gamma_U,\Gamma_\Delta\,\cos(\pi
q/e))$ with $\Gamma_U=0.3\,\Gamma_\Delta=1.5\,E_M$ (see the green curve
in figure~\ref{fig:phasediagram}). The conductance $\bar{G}$ along $c_2$
is shown in figure~\ref{fig:conductance}(b). At $q=0$ the path starts deep
in the band insulator phase and enters the topological phase at $q\approx
0.2$ where the conductance becomes nonzero. When further increasing $q$ we
observe weak oscillations which are due to the finite size lifting $\delta$
of the ground state degeneracy. For larger values of $q$ the conductance
stays zero in the incommensurate and commensurate CDW phases with the
exception of the phase transition between them, where the conductance is
nonzero due to finite-size effects.

Above we have shown how to realize the 1D ANNNI model using Josephson
junction arrays and how to map out its phase diagram by measuring the
tunnelling conductance. In this sense our proposed set-up constitutes a
quantum simulator for the 1D ANNNI model.  In particular, the experimental
control over the system parameters like the gate voltages allows to study
the effects of disorder on the phase diagram.

\section{Relation to nanowires}

Interacting electrons in proximity-coupled
semiconducting nanowires are described by the microscopic Hamiltonian
\cite{stoudenmire:11}
\begin{eqnarray}\label{eq:ham} 
  \hspace{-.7em}H_{\rm\scriptsize NW} &= -  \int\!dx\,\Psi^\dag \left(
  \frac{\partial_x^2}{2m} +\mu + i \alpha \sigma^y \partial_x + E_Z \sigma^z
  \right) \Psi\\ &
  \quad+ \int\!dx\,\Bigl(\Delta_s \Psi_{\uparrow}\Psi_{\downarrow}
  + {\rm H.c.}
  + U_0 |\Psi_{\uparrow}(x)|^2 |\Psi_{\downarrow}(x)|^2\Bigr)
  \nonumber
\end{eqnarray}
with $\Psi(x)= (\Psi_\uparrow(x),\Psi_\downarrow(x))^T$ the electron field
operator, $m$ the electron mass, $\mu$ the chemical potential, $\alpha$
the strength of the spin-orbit coupling, $E_Z$ the Zeeman energy due to the
applied magnetic field, $\Delta_s$ the $s$-wave pairing amplitude, and $U_0$
the (short-range) Coulomb interaction. For sufficiently strong Zeeman energy
(compared to the other energy scales), we only need to consider a single
band similar to the ANNNI model discussed above.  However, projecting the
Hamiltonian \eref{eq:ham} onto a single band strongly reduces the effect
of the interaction. Specifically, we find for the effective interaction
strength in the single-band model $U/E_M = m U_0 \alpha^2/\hbar^2 L E_Z^2
\ll 1$ with $L$ the length of the nanowire. In this way, interacting
nanowires subject to a strong Zeeman field are always in the weak coupling
regime \cite{stoudenmire:11}.  In contrast as we showed above, the strong
interaction regime for spinless fermions is readily accessible in the case
of nanowires in Josephson junction arrays.

\section{Conclusions}

We have analysed a 1D array of Josephson junctions
featuring Majorana modes, where capacitances between adjacent islands lead
to interactions between the Majorana modes. We have shown that repulsive
interactions generically facilitate the topological phase due to their
competition with the on-site charging energies. Finally, we have proposed
a tunnelling experiment to detect the phase boundaries.

\section*{Acknowledgements}

We have benefited from discussions with Eran Sela
and Kirill Shtengel and thank Bernd Braunecker for useful comments. This
work was supported by the Alexander von Humboldt Foundation (FH) and the
DFG through the Emmy-Noether program (DS).

\appendix
\section{Derivation of the quantum phase slip rate}\label{sec:phase_slip}

In this appendix, we present more information about the derivation of
the rates $\Gamma_\Delta$ and $\Gamma_U$. Starting with $\Gamma_\Delta$,
we are interested in the event that the phase $\phi_k$ of a single
island changes by $2\pi$. The relevant tunnelling matrix element
$t_\Delta\propto\langle 2\pi | e^{-i H t} | 0 \rangle$ can be evaluated
in the path-integral formalism by going to imaginary time $\tau=it$,
cf.~\cite{coleman,zinn-justin,altland},
\begin{equation}\label{eq:tunneling}
  t_\Delta \propto \int \mathcal{D} [\phi_k] e^{-\hbar^{-1} \int\!d\tau\,L_E}
\end{equation}
subject to the boundary condition $\phi_k(0)= 0$ and
$\Delta\phi_k=\phi_k(\infty) -\phi_k(0) \in 2\pi \mathbb{Z}\setminus
\{0\}$. To exponential accuracy, the path-integral is dominated by
the classical paths $\phi_k^{{\rm\scriptsize cl},n}$ which minimize the action
$S_E=\hbar^{-1}\int\!d\tau\,L_E$, i.e.,
\begin{equation}\label{eq:semiclassical}
  t_\Delta \sim \sum_{n} e^{- S_E[\phi^{{\rm\scriptsize cl},n}_{k}]}
\end{equation}
where $n$ is an index enumerating the different paths in the case that
there are different minima of the action. 

For $\eta=0$, the relevant part of the action is well-approximated by
(keeping only terms which depend on $\phi_k$)
\begin{equation}\label{eq:action_eta_0}
  S_E[\phi_k]= \int_0^\infty\!d\tau\underbrace{\Bigl[\overbrace{\frac{\hbar}{16
  E_C} \phi_k'^2}^{T}
  - \frac{iq}{2e} \phi_k' + \overbrace{\frac{E_J}{\hbar}
  (1-\cos\phi_k)}^{V_J} \Bigr]}_{L_E},
\end{equation}
where $'$ denotes the derivative with respect to $\tau$ and we have neglected
the potential proportional to $E_M \ll E_J$. As the action does not depend
directly on $\tau$, the energy along the classical path minimizing the action
is conserved,
\begin{equation}\label{eq:energy}
  E= \frac{\partial L_E}{\partial \phi_k'} \phi_k' - L_E = T - V_J.
\end{equation}
For $\tau=0$ we have $T=V_J=0$ such that $E=0$ in our case.

We can express the kinetic energy in terms of the potential and obtain
\begin{equation}\label{eq:kinetik}
  T= \frac{\hbar}{16
  E_C} \phi_k'^2 = E+ V_J
\end{equation}
with which we can get an alternative expression for the measure (the
capacitance matrix acts as a metric)
\begin{equation}\label{eq:measure}
  d \tau = \sqrt{ \frac{\hbar}{16 E_C(E+V_J)} }\, d\phi_k.
\end{equation}
Due to the conservation of energy, we can go over to the Euler-Maupertuis
action $S_0 = S_E+ \int\!d\tau\, E$ (note that in our case $E=0$ such
that $S_0$ is in fact equal to $S_E$) which can be rewritten employing
\eref{eq:measure} as
\begin{equation}\label{eq:s0}
  S_0[\phi_k] = \sqrt{\frac{\hbar}{4 E_C}} \int\!d\phi_k\,\sqrt{E+V_J}
  - \frac{i q}{2e} \Delta \phi_k
\end{equation}
which is independent on imaginary time and only depends on the path chosen
\cite{landau:1}.

The action is minimized for $\Delta \phi_k = \pm 2\pi$ as each additional
phase slip by $2\pi$ increases $S_0$. The expression corresponding to
the first term in \eref{eq:s0} is independent on $\pm$. The classical
path corresponds to increasing $\phi_k$ by $2\pi$ such that, cf~\cite{koch:07},
\begin{eqnarray}\label{eq:s0_classical}
 \sqrt{\frac{\hbar}{4 E_C}} \int\!d\phi_k\,\sqrt{E+V_J}
   = \sqrt{\frac{E_J}{4 E_C}}
  \int_0^{2\pi} \!d\phi_k\, (1-\cos \phi_k) = \sqrt{\frac{8 E_J}{E_C}}. 
\end{eqnarray}
We obtain the final result
\begin{eqnarray}\label{eq:result}
 t_\Delta &\sim e^{-\sqrt{8 E_J/E_C}} \sum_{\Delta\phi_k=\pm 2\pi}
 e^{-iq \Delta \phi_k/2e} 
 \sim   e^{-\sqrt{8 E_J/E_C}}  \cos(q \pi/e).
\end{eqnarray}
valid up to exponential accuracy. In the main text, we use the result
$t_\Delta = \Gamma_\Delta \cos(\pi q/e)$ with $\Gamma_\Delta \simeq E_C^{1/4}
E_J^{3/4} e^{-\sqrt{8 E_J/E_C}}$.  In fact the charge $q$ should be replace
by $q + e n_k$. The reason is that due to the Majorana term the action is
not $2\pi$ but only $4\pi$ periodic in $\phi_k$. In the calculation above,
we however assume the action to be $2\pi$ periodic. In fact, the action
can be made $2\pi$ periodic by a gauge transformation on the expense of
replacing $q\mapsto q+ e n_k$. More information on this rather subtle point
can be found in \cite{fu:10,heck:11}.

The prefactor $E_C^{1/4} E_J^{3/4}$ of $\Gamma_\Delta$ depends on the
shape of the potential close to the turning points $\phi_k \approx 0,\pm
2\pi$ and cannot be obtain in our simple semiclassical analysis which only
captures the physics up to exponential accuracy. However, the scaling of
the prefactor can be obtained from summing up the instanton contributions
\cite{zinn-justin} or by matching it to the exact solution of the Mathieu
equation \cite{koch:07}. In our case, the potential always is given by
$V_J \simeq \frac12 E_J \phi_k^2$ for $\phi_k \ll 1$ such that the same
prefactor $E_C^{1/4} E_J^{3/4}$ (from the Mathieu equation) appears for
all the tunnelling amplitudes.

In the case $\eta\neq0$, it is important to notice that the phases on the
different islands do not completely decouple. In lowest order in $\eta$, we
need to take the phases $\phi_{k\pm1}$ on the islands $k\pm1$ into account.
Due to the symmetry of the problem, we know that $\phi_{k-1} (\tau) =
\phi_{k+1}(\tau)$. The relevant part of the action reads
\begin{eqnarray}\label{eq:action_eta}
  S_E &= \int_0^\infty d\tau\Bigl[ \frac{\hbar }{16E_C} \bigl(\phi_k'^2 
  + 2 (1-\eta)\phi_{k+1}'^2 \bigr) \nonumber
  \\&\qquad\qquad-\frac{i q}{2e} (\phi_k' + 2\phi_{k+1}')
  + \underbrace{\frac{E_J}{\hbar}(3 - \cos \phi_k -2 \cos \phi_{k+1})}_{V_J} 
  \Bigr]. 
\end{eqnarray}
Following the same line of calculation as going from \eref{eq:action_eta_0}
to \eref{eq:s0}, we obtain
\begin{equation}\label{eq:s02}
  S_0 = \sqrt{\frac{\hbar}{4 E_C} } \!
  \int\!d\phi_k  \underbrace{\sqrt{[1 + 2 (1-\eta)
  \phi_{k+1}'(\phi_k)^2] V_J}}_{\mathcal{L}_{\rm\scriptsize eff}} 
  - \frac{i q \Delta\phi_k}{2e} ,
\end{equation}
where we expressed the path by giving $\phi_{k+1}$ as a function of
$\phi_k$.  As we are interested in processes where $\phi_k$ changes by
$\Delta\phi_k=\pm 2\pi$, we need to find $\phi_{k+1}(\phi_k)\colon [0,\pm
2\pi] \mapsto \mathbb{R}$ with $\phi_{k+1}(0) = \phi_{k+1}(\pm 2\pi)$ such
that the action is minimized.  We will find the solution which corresponds
to $\Delta \phi_k = 2\pi$ below. The second solution with $\Delta \phi_k =
-2\pi$ can by obtained via the symmetry $\phi_i \mapsto - \phi_i, \forall i$
of the Lagrangian.

As the second term in \eref{eq:action_eta} is independent of the path
(it depends only on the boundary condition), we only need to minimize the
first term. The extremum is attained when the function $\phi_{k+1}(\phi_k)$
fulfils the Euler-Lagrange equation
\begin{equation}\label{eq:el_equation}
  \frac{d}{d\phi_k} 
  \frac{\partial \mathcal{L}_{\rm\scriptsize eff}}{\partial \phi_{k+1}'}
  = \frac{\partial \mathcal{L}_{\rm\scriptsize eff}}{\partial \phi_k}.
\end{equation}
To first order in $\eta$ [assuming $\phi_{k+1} \in \Or(\eta)$], the equation
assumes the form
\[
  4 (1-\cos \phi_k)\phi_{k+1}'' +2 \sin (\phi_k) \, \phi_{k+1}' - 
  \eta \sin \phi_k -2
  \phi_{k+1} =0.
\]
Employing the substitution $\phi_{k+1}(\phi_k) = \tan(\phi_k/4) f(\phi_k)$
reduces the equation to first order equation in $f'$ of the form
\begin{eqnarray}\label{eq:first_order}
  2 \cos (\phi_k/4) [8 \sin^2(\phi_k/4) f''(\phi_k) 
 -\eta \cos (\phi_k/2)  ] \nonumber\\
  \qquad\qquad + 2 [3 \sin(\phi_k/4) + \sin(3 \phi_k/4)] f'(\phi_k) =0
\end{eqnarray}
which can be integrated with the solution
\begin{equation}\label{eq:sol}
  f(\phi_k) = \eta \log\left[\frac12 \sin (\frac12\phi_k)\right] - 
  \eta \frac{\log \cos (\frac14 \phi_k)}{
  \sin^{2}(\frac14
  \phi_k)}.
\end{equation}
Plugging the solution into \eref{eq:s02} and retaining the first
nonvanishing term in $\eta$ yields
\begin{equation}\label{eq:s0_solutin}
  S_0 = \sqrt{8 E_J/E_C} \left[ 1 +
  \frac{\pi^2-12}{96} \eta^2 + \Or(\eta^4) \right] - \frac{i q \Delta
  \phi_k}{2e}.
\end{equation}
Summing up the two contributions with $\Delta \phi_k = \pm 2\pi$, we obtain a
term proportional to $\cos (\pi q/e)$ as before, with the proportionality
constant given by $\Gamma_\Delta = E_C^{1/4} E_J^{3/4} e^{- \sqrt{8 E_J/E_C}
[
1 + (\pi^2-12)\eta^2/96]}$.

For $\eta \neq 0$, we get additionally a next-nearest neighbour interaction
due to phase slips where both $\phi_k$ and $\phi_{k+1}$ change by $2\pi$. In
fact, due to the symmetry of the problem, we can set $\phi_{k+1}(\tau) =
\phi_k(\tau)$. The term of the action which change with $\phi_k$ and
$\phi_{k+1}$ are given by \eref{eq:action_eta_0} for each of the islands and
the additional contribution of the cross capacitance. Thus, we have
\begin{equation}\label{eq:se3}
  S_E = \int_0^\infty\!d\tau\, \biggl[ \frac{\hbar (2 - \eta)}{16 E_C} 
  \phi_k'^2  + 2 E_J (1- \cos \phi_k) \biggr]
\end{equation}
which leads to
\begin{eqnarray}\label{eq:s03}
  S_0 &= \sqrt{ \frac{(2-\eta) E_J}{2 E_C} } \int\!d\phi_k\, \sqrt{ 1-\cos
  \phi_k} - \frac{i q (\Delta \phi_k + \Delta \phi_{k+1})}{2e}    \nonumber\\
   &= 4 \sqrt{(2 -\eta) E_J/E_C}- \frac{i q (\Delta \phi_k + \Delta
 \phi_{k+1})}{2e} .
\end{eqnarray} 
As $\Delta \phi_{k+1} = \Delta \phi_k$ (because the two phases slip together),
the tunnelling amplitude thus assumes the form $\Gamma_U \cos ( 2\pi q/e)$ with
$\Gamma_U = E_C^{1/4} E_J^{3/4} e^{- \sqrt{16(2-\eta) E_J/E_C}}$.

\end{document}